\documentclass[12pt,preprint]{aastex}

\newcommand{\degree}{$^{\rm{o}}$}       
\newcommand{\rosat}{{\it{ROSAT}}}       

\newcommand{\hone}{H~{\footnotesize{I}}}  	
\newcommand{\halpha}{H{${\alpha}$}}		
\newcommand{\hbeta}{H$\beta$}  		
\newcommand{\nittwo}{[N$_{\rm{II}}$]}  
\newcommand{\oxyone}{[O{\footnotesize{I]}}}	
\newcommand{\oxythree}{[O$_{\rm{III}}$]} 
\newcommand{\sultwo}{[S~{\footnotesize{II]}}}	
\newcommand{\sultwoa}{[S~{\footnotesize{II}}]\ (6717\ \AA)} 
\newcommand{\sultwob}{[S~{\footnotesize{II}}]\ (6731\ \AA)} 

\newcommand{\gsnr}{G65.2$+$5.7}			

\shorttitle{G65.2+5.7}
\shortauthors{Shelton, Kuntz, and Petre}
\begin{document}

\title{G65.2+5.7: A Thermal Composite Supernova Remnant with a
Cool Shell}


\author{R. L. Shelton\altaffilmark{1},
K. D. Kuntz\altaffilmark{2},
R. Petre\altaffilmark{3}}
\affil{$^1$ The Department of Physics and Astronomy and the 
Center for Simulational Physics at the University of Georgia,
Athens, GA 30602}
\affil{$^2$The Department of Physics at the University 
University of Maryland at Baltimore County}
\affil{$^3$NASA's Goddard Space Flight Center}
\email{rls@hal.physast.uga.edu}
\email{kuntz@milkyway.gsfc.nasa.gov}
\email{rob@milkyway.gsfc.nasa.gov}

\begin{abstract}

This paper presents archival \rosat\ PSPC observations of the 
\gsnr\ supernova remnant (also known as G65.3+5.7).  
Little material obscures this remnant and so it was
well observed, even at the softest end of 
\rosat's bandpass ($\sim0.11$ to 0.28~keV).
These soft X-ray images reveal the remnant's
centrally-filled morphology which, in combination with existing
radio frequency observations, places \gsnr\ in the thermal composite 
(mixed morphology) class of supernova remnants.
Not only might \gsnr\ be the oldest known thermal composite supernova
remnant, but
owing to its optically revealed cool, dense shell, this remnant
supports the proposal that thermal composite supernova remnants lack
X-ray bright shells because they have evolved beyond the adiabatic phase.
These observations also reveal a slightly extended point source
centered on RA = 19$^{\rm{h}}$ 36$^{\rm{m}}$ 46$^{\rm{s}}$, 
dec = 30$^{\rm{o}}$ 40' 07'' and extending 6.5 arcmin in radius
in the band 67 map. The source of this emission has yet to be
discovered, as there is no known pulsar at this location.
\\

\end{abstract}




\section{Introduction}

	Within a decade of being identified as a 
supernova remnant (SNR), \gsnr\ (also called 
G65.3+5.7) was the subject of several 
observational papers published in quick succession.  
However, once its radio and optical images and its
optical and X-ray spectra were recorded, the remnant was
ignored.  Now, a quarter of a century later, \gsnr\ has 
once again become interesting, this time because it helps to explain 
an unusual class of supernova remnants called 
thermal composite SNRs (also known as mixed morphology SNRs).

Both of these terms
describe the combination of a shell-like radio continuum
morphology with a centrally-bright X-ray morphology in
which the X-ray emission is due to thermal, rather than synchrotron,
processes.
Compared with shell-type and plerionic SNRs, these remnants
have more mysterious origins.
Various workers have suggested that 
thermal composite SNRs lack X-ray shells
because they have evolved into the radiative phase (and thus
have cool, X-ray dim shells), or that 
their centers are X-ray bright because thermal conduction, other 
forms of entropy mixing, or cloudlet evaporation has 
enhanced the central densities.  Ejecta enrichment or
dust destruction could increase the metallicities in the centers,
also enhancing the X-ray emission.
Alternatively, a thermal composite morphology may be due to
collisions with molecular clouds.
Interested readers are directed to
\citet{white-long}, \citet{long-etal}, 
\citet{cox-etal-99}, \citet{shelton-etal-99}, \citet{chevalier}, 
\citet{yokogawa-etal}, \citet{kawasaki}, and \citet{shelton-kuntz-petre} 
for additional discussions of these processes.

Currently, there are few observations of the ``smoking gun''
for the radiative phase evolution hypothesis: \hone\ shells.  
The \hone\ shell on W44 is the only possible example
(\citet{koo-heiles}, with interpretations in
\citet{cox-etal-99} and \citet{shelton-etal-99}).
Here we demonstrate that another 
thermal composite supernova remnant,
\gsnr, is clearly in the radiative phase.
Ironically, this demonstration is not 
performed by discovering that a previously known
thermal composite SNR is in the radiative phase, but is done by demonstrating
that a previously known radiative phase SNR is a thermal composite.
In Section 2, we discuss published optical observations, indicating
that \gsnr\ has entered the shell formation phase;  
we also discuss published radio and
low spatial resolution X-ray observations, showing that \gsnr\ has a
shell-like radio continuum morphology and emits thermal X-rays. 
With these characteristics, \gsnr\ meets
two of the three requirements for the thermal composite classification.
The third requirement, that the remnant has a centrally
filled X-ray morphology, is established in Section 3,
through the analysis of archival \rosat\ observations.
Furthermore, model fits to the \rosat\ data imply that the 
temperature peaks in
the center and decreases mildly with radius, as is found for 
other thermal composite supernova remnants.
In Section 4, we summarize the results.\\

\section{Previous Observations}

	The first observations of \gsnr, those made with the
red prints of the Palomar Sky Survey \citep{sharpless}, 
only found two bright filaments, S91 and S94, but failed to
recognize the entire outline of the remnant.  The true extent
of the remnant remained unknown until \citet{gull-etal}'s
optical emission line survey of the Galactic plane 
revealed filaments circumscribing a 
4\degree $\times$ 3.3\degree\ ellipse.  
\citet{reich-etal}'s radio frequency data also suggested a
similarly sized shell-type remnant.

	Even before its true extent was revealed, 
\citet{sabbadin-odorico} used their \halpha, \nittwo, \sultwoa, and
\sultwob\ measurements to show that the S91 and S94
filaments were dense ($n > 450$ cm$^{-3}$) shell fragments 
belonging to a large supernova remnant.
\citet{fesen-etal-85} made additional optical emission line
measurements of the S91 filament and of one of the brightest
\oxythree\ filaments.
Both regions have large \oxyone/\hbeta\ ratios, which,
when compared with the shock models of \citet{raymond-etal}, indicate that
the photons were emitted by thick ($N_H > 10^{18.3}$ cm$^{-2}$)
postshock cooling regions behind moderate velocity 
($v_s = 100$ to 120 km s$^{-1}$) shocks.
Given its \oxyone/\hbeta\ ratio,
S91's relatively modest \oxythree/\hbeta\ ratio indicates
that the cooling and recombination behind the shock is ``complete''
\citep{raymond-etal}.  
The cooling zone along \citet{fesen-etal-85}'s second
pointing direction appears to be well developed, but incomplete.
Given that the second pointing direction was chosen for its
anomalously large \oxythree/\hbeta\ ratio, this portion 
may be less complete than most of the remnant's shell.
Similarly, \citet{mavromatakis-etal} found regions of incomplete 
cooling as well as regions of nearly complete cooling.
The shell temperature found from the older \oxythree\ and \sultwo\
measurements is $\sim$38,000~K 
(Fesen, Blair \& Kirshner 1985; Sabbadin \& D'Odorico 1976).
\citet{mavromatakis-etal} calculated shock velocities of
90 to 140 km sec$^{-1}$, which are roughly consistent with 
\citet{rosado}'s kinematically determined expansion velocity
of $90 \pm 30$ km s$^{-1}$.    
%
In summary, \gsnr\ is sufficiently evolved for the shock to be 
weak and the gas behind it to be relatively cool and recombined.

Multiple techniques have been employed to estimate the remnant's
size and distance.  \citet{reich-etal}
applied the $\Sigma$-D relation to their 1420 MHz data
obtaining a diameter of $75^{+50}_{-25}$ pc
and a distance of $900^{+600}_{-300}$ pc.
\citet{lozinskaya} applied the Galactic kinematic relationship
to her Fabry-Perot interferometric \halpha\ data in order to 
estimate the distance as 800 pc and the minor and major axes of
this elliptically shaped remnant as 56 and 64 pc, respectively.
With a distance of $900^{+600}_{-300}$ pc and Galactic 
latitude of 5.7\degree, the remnant lies 
$90^{+60}_{-30}$~pc above the Galactic disk.  
Furthermore, \gsnr's 
age has been estimated as $\sim 3 \times 10^5$~yr
(\citet{gull-etal}, using radiative phase SNR relations).\\



While others were observing the remnant with optical and radio-frequency
detectors, \citet{snyder-etal} and \citet{mason-etal} were observing it
with X-ray detectors.
\citet{mason-etal}'s {\it{HEAO~1}} (bandpass = 0.2 to 2.5 keV, 
FWHM = 3\degree) spectrum exhibits line
emission, the hallmark of thermal emission from hot plasma.
The model which best fits the {\it{HEAO~1}} spectrum is 
a Raymond and Smith thermal model with a temperature between
$2.0$ and $3.6 \times 10^6$~K 
\citep{mason-etal}. 
Although they were
not able to resolve the emission spatially with the
{\it{HEAO\ 1}} data, they were able to show that the
emission region is not point-like.
In the following section, we extend upon this work.  We
use archival \rosat\ data to map the remnant, confirming that
the emitting gas is extended and finding that it fills the
remnant's outline, lacks a bright shell, and exhibits a slowly 
decreasing temperature gradient.   

\section{ROSAT PSPC X-Ray Observations} 

\begin{figure}           
  \plotone{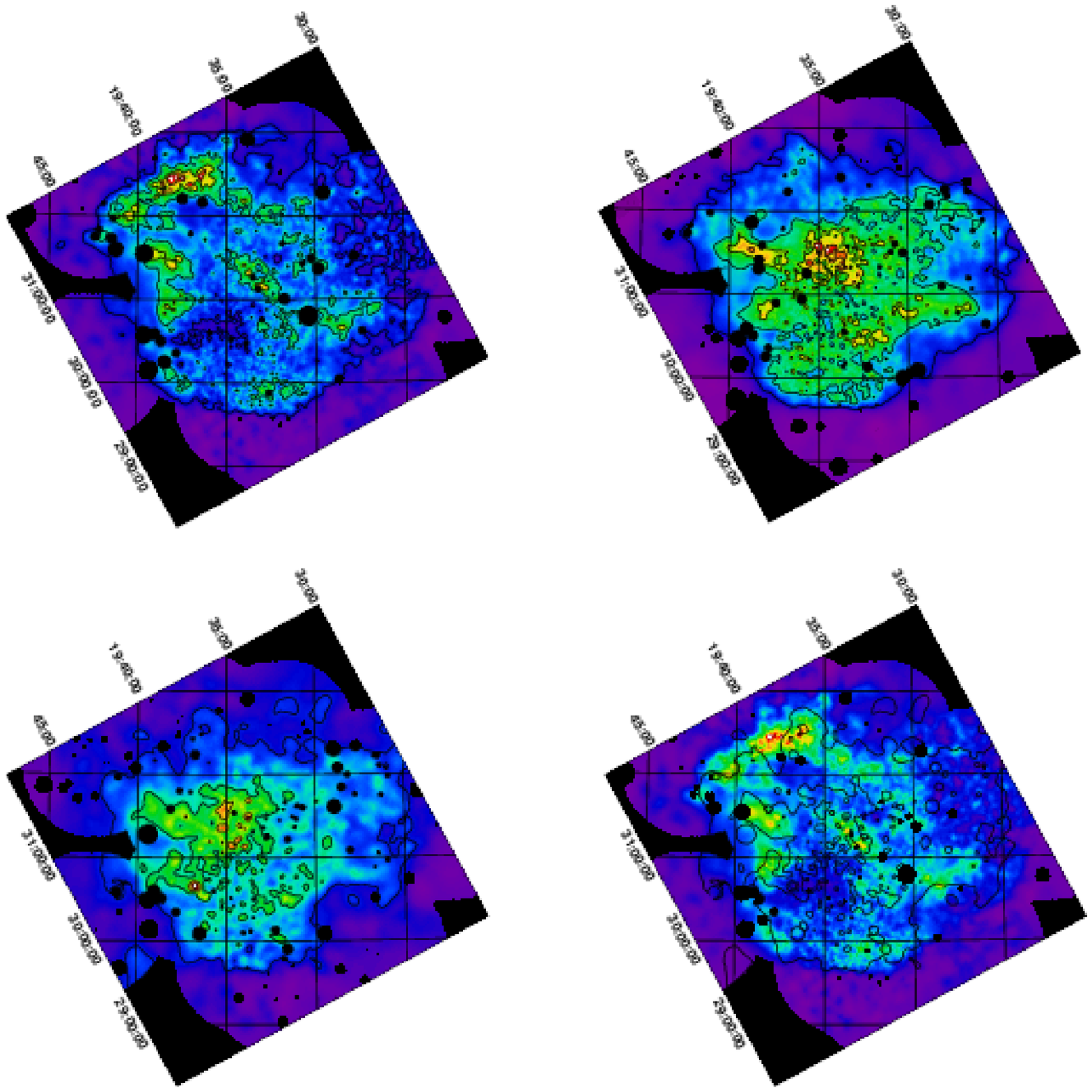}
  \caption{Mosaic maps of G65.2+5.7 in J2000 coordinates.  The
black disks show where bright point sources have been removed.
{\it{Top Left}}: Map in the \rosat\ 1L2 band, made from PI channels 11 to 41 
and sensitive to 0.11 to 0.284 keV radiation.  
Contours are drawn
at 125, 250, 375, and 500 $\times 10^{-6}$ counts s$^{-1}$ arcmin$^{-2}$.
{\it{Tob Right}}: Map in the \rosat\ 45 band, made from PI channels 52 to 90 
and sensitive to 0.44 to 1.21 keV radiation.  Contours are drawn
at 75, 150, 225, and 300 $\times 10^{-6}$ counts s$^{-1}$ arcmin$^{-2}$.
{\it{Bottom Left}}: Map in the \rosat\ 67 band, made from PI channels 91 to 201 
and sensitive to 0.73 to 2.04 keV radiation. Contours are drawn
at 50, 100, 150, and 200 $\times 10^{-6}$ counts s$^{-1}$ arcmin$^{-2}$.
{\it{Bottom Right}}: Map in the \rosat\ 1L2 band, overlayed with band 67 contours.
The semi-circle contour in the lower left of the image is an artifact
of the pointsource removal in the band 67 map.
These X-ray images reveal \gsnr's ``center filled'' morphology as
well as its slight spectral hardness gradient.
}
\label{rosatimages.fig}
\end{figure}

Between August 1995 and September 1997, the remnant was
mapped with a series of overlapping \rosat\ PSPC observations.
We processed the \rosat\ PSPC data
(removed the contamination, scattered solar X-rays, 
long-term enhancements, after-pulses, 
particle background, and bright point sources)
and mosaiced the pointings
according to the extended object analysis
procedures described in \citet{snowden-kuntz} and \citet{kuntz-snowden}.
The resulting images in the
\rosat\ 1L2 band (sensitive to 0.11 to 0.284 keV photons),
45 band (sensitive to 0.44 to 1.21 keV photons), and
67 band (sensitive to 0.73 to 2.04 keV photons) are
displayed in Figure~\ref{rosatimages.fig}, along with 
a 1L2 band image overlayed with contours from the 67 band map.
With the exception of a ridge of band 1L2 emission along the northeast
edge and roughly coincident with bright optical filaments,
the periphery is X-ray dim.
Most of the projected interior shows X-ray emission.  
Because of its generally ``centrally filled'' X-ray
morphology, thermal X-ray emission, and 
``shell-type'' radio continuum morphology, the remnant should
now be classified as a thermal composite.

Comparison of the soft and hard band images shows that the X-ray
emission originating in the SNR interior is
slightly harder than that originating nearer to the edges.  
We quantify this trend with a spectral model.
The gas is assumed to be near collisional ionizational equilibrium,
as is observed both in similar remnants
such as W44 
(Harrus et al. 1997; Shelton, Kuntz \& Petre 2004), 
3C391 (Chen \& Slane 2001), and 0045-734 in the SMC (Yokogawa et al 2002),
and in simulated remnants which have evolved long enough
to have centrally-peaked X-ray morphologies \citep{shelton1999}.
Given \rosat's spectral resolution, the differences between
equilibrium and nearly equilibrium spectra are inconsequential,
as are differences between newer and older versions of the 
\citet{raymond-smith} spectral code.
In general, the model spectra are calculated and handled according to 
the techniques described
in \citet{kuntz-snowden-2000}.  Thus, we used the 
\citet{raymond-smith} spectral code to create a suite of
collisional ionizational equilibrium spectral models.
The electron temperature ($T_e$) ranged from 
$10^5$~K to $10^8$~K, with increments
in log($T_e$) of 0.05.
In order to account for
the absorption by the interstellar medium, we multiplied our
suite of model spectra by $e^{-\tau}$, where $\tau$ is the product
of the absorption column density ($N_H$) and the absorption cross
section \citep{morrison-mccammon}.  $N_H$ ranged from
$10^{16}$ cm$^{-2}$ to $10^{24}$ cm$^{-2}$, with increments
in log($N_H$) of 0.10.  The resulting spectra were
convolved with the \rosat\ PSPC response matrix, yielding
the countrates in each \rosat\ PSPC band.

\begin{figure}           
  \plotone{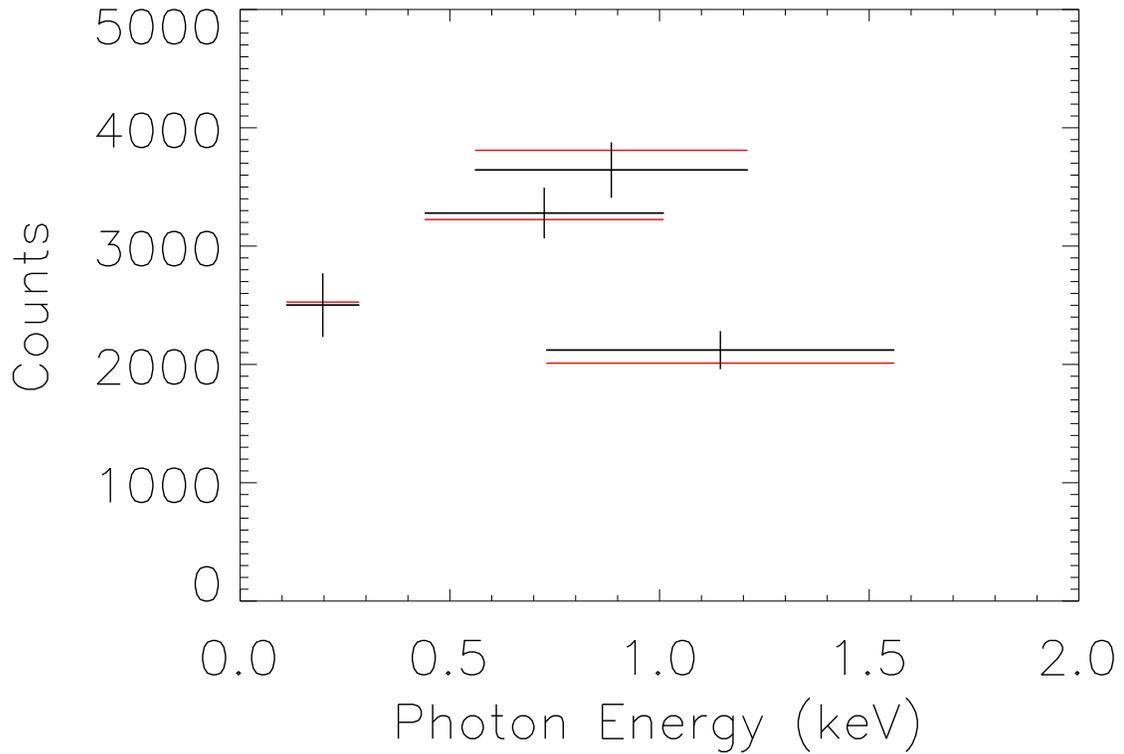}
  \caption[Spectrum]
{Observed (black) and model (red) spectra for the seventh annulus.
The 1$\sigma$ error bars were calculated from the pixel to 
pixel variation in the observational data.  
Band 2 data have been ignored owing to complications
with an absorption edge in the detector, while band 7 data
have been ignored owing to the low countrate.  
The \rosat\ PSPC band photon energies were taken from \citet{snowden-etal}.
This fit has $\chi^2 = 1.1$ with 1 degree of freedom and is neither
the best nor the worst case among the annular fits. }
\label{spectrum.fig}
\end{figure}

We prepared the data for comparison with the calculated models by
excluding counts from regions of the X-ray image that were contaminated by
bright point sources or dominated by the
Galactic background emission, dividing the
the remaining SNR area into ten, 
concentric, nearly equal area annuli.
These annuli 
were centered on 
RA = 19$^{\rm{h}}$ 33$^{\rm{m}}$ 41$^{\rm{s}}$, 
dec = 31$^{\rm{o}}$ 15' 57'' in J2000 coordinates and had outer radii of 
26, 39, 48, 56, 64, 71, 79, 86, 94, and 122 arcmin, respectively.  
Regions beyond the SNR and regions contaminated by point sources
(including the slightly extended hard source centered on
RA = 19$^{\rm{h}}$ 36$^{\rm{m}}$ 46$^{\rm{s}}$, 
dec = 30$^{\rm{o}}$ 40' 07'')
were excluded from the annuli areas.  
As a result, in order to create approximately equal area masks,
the outermost annuli were extended to relatively large radii.
We then calculated the countrates in each of the \rosat\ PSPC bands
for each annulus.
The background countrate for each
band was calculated from the cleaned data for the portion of the 
observed sky that lies outside the SNR's footprint and subtracted from
the countrate for each annulus, resulting in the ``observed countrates''.
The standard chi-squared test was used to compare the
observed countrates in the \rosat\ 1L, 4, 5, and 6 bands for each annulus
with the countrates predicted for the spectral models.  
Band 2 was omitted, owing to
complications of an absorption edge in the detector and band 7
was omitted owing to its low countrate.  The best fitting models were
then selected for each annulus.
Figure~\ref{spectrum.fig} displays
the observed and model spectra for the seventh annulus (spanning
the region between 79 and 86 arcmin from the remnant's center).

\begin{figure}           
  \plotone{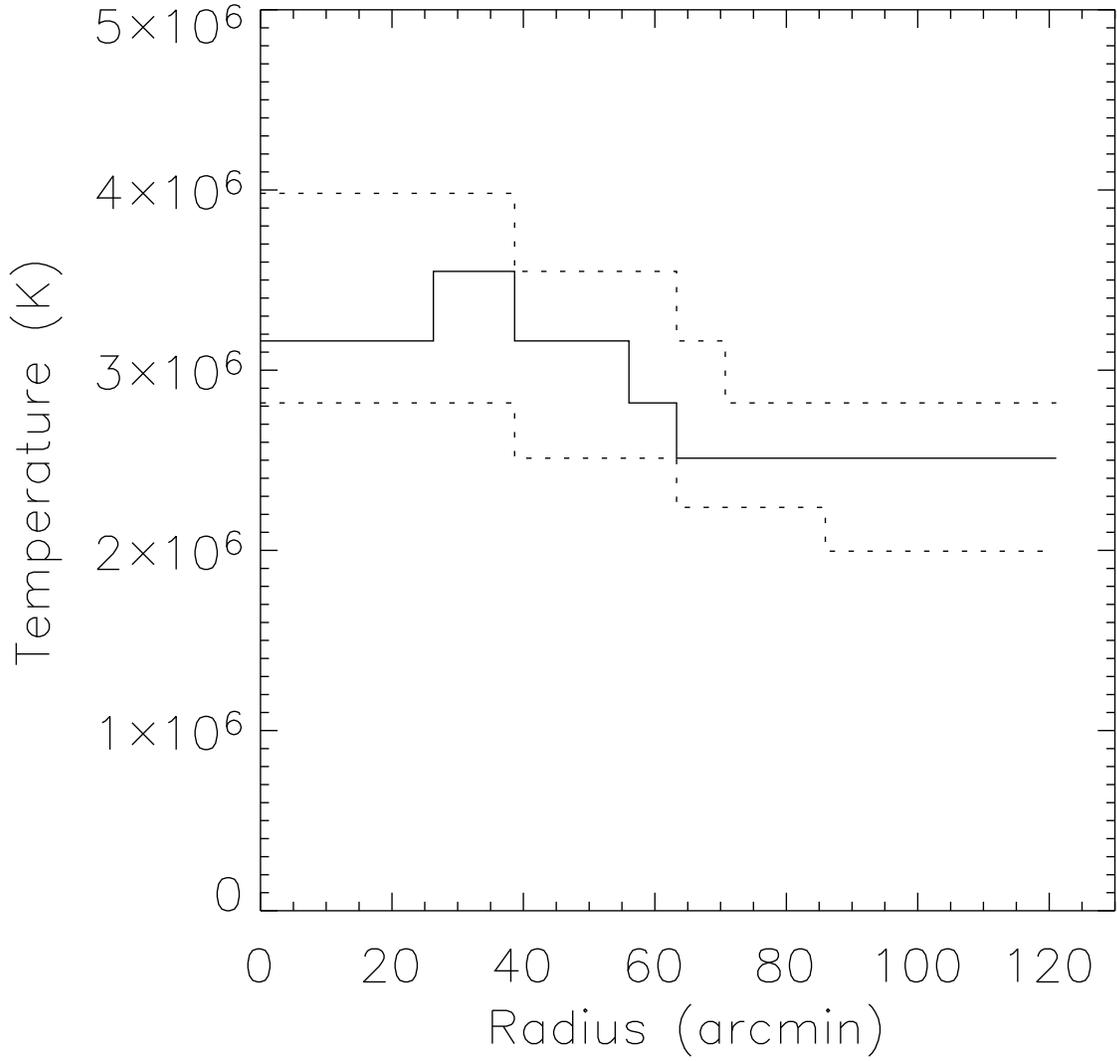}
  \caption[Temperature versus Radius for best fitting models]
{The solid histogram traces the electron temperature of the 
best fitting plasma model as a function of radius.  The dotted histograms 
trace the $\pm1\sigma$ values.  }
\label{temp.fig}
\end{figure}

The best-fit temperatures range from 2.5 to $3.5 \times 10^6$~K and 
decrease slightly with radius 
(see Figure~\ref{temp.fig}).  
The shallow temperature gradient
is a common feature of thermal composite SNRs 
(i.e., 3C400.2 \citep{yoshita-etal}, Kes 27 \citep{enoguchi-etal}, 
W44 \citep{shelton-kuntz-petre}).
In the best fit models for the various regions, the
column densities ranged 
from $8 \times 10^{19}$ to $1.3 \times 10^{20}$~cm$^{-2}$, 
but did not exhibit a radial trend.

The SNR's apparent flux of 0.11 to 2.04 keV photons is 
$6.8 \times 10^{-11}$ ergs cm$^{-2}$ s$^{-1}$.
If there had been no interstellar
scattering, the flux would have been 
$8.4 \times 10^{-11}$ ergs cm$^{-2}$ s$^{-1}$.
Assuming that the remnant is 900 pc distant, its luminosity
is $8.2 \times 10^{33}$ ergs s$^{-1}$ and the electron density
of the X-ray emitting gas is $0.013$ cm$^{-3}$.  
Based on the theoretical predictions and hydrocode simulations
of the archetypical thermal composite supernova remnant, W44
(\citet{cox-etal-99}, \citet{shelton-etal-99}), we expect the ambient
density to be about 10 times the density within the remnant's
center.  Thus, the ambient density would be roughly 0.1 cm$^{-3}$,
a reasonable value at \gsnr's location 90 pc above the Galactic
plane.   The electron density is lower than that of W44, causing
the luminosity to be lower as well.

In addition to displaying a centrally concentrated SNR morphology,
the band 45 and band 67 X-ray images also reveal a 
``slightly extended point source''
centered on RA = 19$^{\rm{h}}$ 36$^{\rm{m}}$ 46$^{\rm{s}}$, 
dec = 30$^{\rm{o}}$ 40' 07'' and extending approximately 6.5 arcmin
in radius in the harder band map and less far in the band 45 map.
The 1L2 band map does not show excess emission at or near this location.
The observed spectrum is too noisy for meaningful fitting to model spectra
and there is no known pulsar at this location.
Although the Princeton-Arecibo survey for millisecond pulsar's
did not find a pulsar coincident with the observed bright spot, 
it did find a pulsar, J1931+30, within the remnant's outline
\citep{camilo-etal}. 
We did not find excess X-ray emission at its location, 
RA = 19$^{\rm{h}}$ 31$^{\rm{m}}$ 28$^{\rm{s}}$, 
dec = 30$^{\rm{o}}$ 35'.

\section{Conclusion} 

These \rosat\ observations, in combination with existing 
radio observations, establish \gsnr's membership in the 
thermal composite (mixed morphology) class of supernova remnants.
The radio continuum shells and thermal X-ray bright centers of
this puzzling class of remnants have inspired several theoretical
explanations.  In one scenario, these remnants have evolved beyond
the adiabatic phase into the radiative phase.  As a result, their
peripheries are now cool and not X-ray emissive.  This scenario
can explain why these remnants lack X-ray bright shells, while
other explanations are required in order to account for
the extreme brightnesses of these remnants' centers.
Here, we note that crucial support for the evolutionary hypothesis
is provided by optical evidence of 
nearly complete cooling behind \gsnr's relatively slow shockfront.
\\

If \citet{gull-etal}'s age estimate is correct, 
then \gsnr\ is the oldest known thermal composite supernova remnant.
The fact that several other thermal composite supernova remnants 
neighbor molecular clouds has led to the suggestion that
collisions with molecular clouds cause the characteristic 
thermal composite morphology.  In order to test this
suggestion on \gsnr, we searched the \citet{dame-etal-2001}
CO survey data and the \citet{hartmann-burton} \hone\ survey data
near $l = 65.2^{\rm{o}}, b = 5.7^{\rm{o}}$ at the remnant's systemic
velocity ($+8 \pm 10$~km s$^{-1}$, \citet{lozinskaya}).  
Confusion with Galactic material made the search inconclusive.
The possibility of a SNR-cloud interaction could potentially be
addressed with a targeted search for CO emission features 
outlining the remnant.\\

\vspace{1cm}

\noindent{\bf{Acknowledgements}}

R.L.S. wishes to thank the 
National Research Council and NASA's Long Term Space Astrophysics
Grant (NAG5-10807) for financial support and the anonomous referee
for his/her helpful comments.
K.D.K. wishes to thank GSRP for financial support.
This research has made use of data obtained from the High Energy 
Astrophysics Science Archive Research Center
(HEASARC), provided by NASA's Goddard Space Flight Center.

\clearpage

\pagebreak

\pagebreak

\pagebreak

\end{document}